\documentclass[pra,aps,psfig,epsf,amsmath,amssymb,groupedaddress]{revtex4}
\usepackage{graphicx}
\usepackage{psfrag}
\usepackage{amsmath,amssymb}
\usepackage{colordvi}
\usepackage{array}

\renewcommand{\b}{\beta}

\newcommand{\D}{\Delta}

\renewcommand{\o}{\omega}

\newcommand{\lb}{\label}

\newcommand{\PD}[2]{\frac{\partial{#1}}{\partial{#2}}}

\newcommand{\BK}[1]{\left[#1\right]}

\newcommand{\ket}[1]{|{#1}\rangle}

\newcommand{\be}{\begin{equation}}
\newcommand{\ee}{\end{equation}}
\newcommand{\ba}{\begin{eqnarray}}
\newcommand{\ea}{\end{eqnarray}}

\begin{document}

\title{Atom Loss Maximum in Ultra-cold Fermi Gases}
\author{Shizhong Zhang and Tin-Lun Ho}
\affiliation{Department of Physics, The Ohio-State University, Columbus, Ohio 43210}
\date{\today}
\begin{abstract}

Recent experiments on atom loss in ultra-cold Fermi gases all show a maximum at a magnetic field below Feshbach resonance, where the s-wave scattering length is large (close to inter-particle distance) and positive.  These experiments have been performed over a wide range of conditions, with temperatures and trap depths spanning over three decades. Different groups have come up with different explanations, among them the emergence of Stoner ferromagnetism.  Here, we show that this maximum is a consequence of two major steps.  The first is the establishment of a population of shallow dimers, which is the combined effect of dimer formation through three-body recombination,  and the dissociation of shallow dimers back to atoms through collisions.   The dissociation process will be temperature dependent, and is affected by Pauli blocking at low temperatures.
The second is the relaxation of shallow dimers into tightly bound dimers through atom-dimer and dimer-dimer collisions. 
 We have constructed a simple set of rate equations describing these processes. Remarkably, even with only a few parameters, these equations reproduce the loss rate observed in all recent experiments, despite their widely different experimental conditions.  Our studies show that the location of the maximum loss rate depends crucially on experimental parameters such as trap depth and temperature.  These extrinsic characters show that this maximum is not a reliable probe of the nature of the underlying quantum states. The physics of our equations also explains some general trends found in current experiments. 

\end{abstract}
\maketitle

\section{Introduction}
Three-body processes occupy a unique place in the study of ultra-cold atomic gases. On the one hand, while two-body elastic collisions are important for  thermal equilibrium, three-body processes are crucial for producing molecules of interested, such as those in the BEC-BCS crossover. On the other hand, three-body processes can also lead to atom loss from the trap, which limits the lifetime of the sample and constraints the time scale over which experiments can be performed. In addition, three-body process can also be used as diagnostic tool.  Recent advances in Efimov physics are resulted from using three-body loss to identify the presence of trimers in a bosonic gas of $^7$Li\cite{Kraemer2006,Pollack2009}.\\

Theoretically, the study of three-body problem dates back to the early days of quantum mechanics and has revealed many exciting new phenomena, such as Thomas effect \cite{Thomas1935} and Efimov physics \cite{Braaten2006}. In the case of interest here, namely, 
an equal population of two hyperfine states of  fermionic quantum gas such as $^6$Li,  Efimov physics is irrelevant. What is important is the process of three-body recombination, in which three atoms collide to form a shallow dimer,  with the third atom carrying away the energy released in the process.  We shall refer this as Process ({\bf I}) and denote it as $A_{\uparrow}+A_{\downarrow}+A_{\uparrow} \to D + A_{\uparrow}$,  and
$A_{\downarrow}+A_{\uparrow}+A_{\downarrow} \to D + A_{\downarrow}$, where $D$ is the shallow dimer;  $A_{\uparrow}$ and $A_{\downarrow}$ are the atoms in different hyperfine states which we simply denote as $\uparrow$ and $\downarrow$. 
Note that process ({\bf I}) only occurs when the two-body s-wave scattering length $a_s$ is positive, since bound states only exist for 
$a_{s}>0$. 
 The reverse of Process ${\bf (I)}$ is the dissociation of shallow dimers back to atoms, 
 $A_\uparrow+D\to A_\uparrow+A_\uparrow+A_\downarrow$, which we refer to as Process ${\bf (I')}$.  The combined effect of ${\bf (I)}$ and ${\bf (I')}$ is to provide 
 a net population of shallow dimers.  \\

  Once shallow dimers $D$ are formed, they can further relax into deep bound states in the van der Waals potential through either atom-dimer collision, 
 $A+D \to D^*+A$ (referred to as process ({\bf II})), or dimer-dimer collision, $D+D\to D^*+D^*$ (referred to as process ({\bf III})), where $D^*$ is the deep bound state.  These two processes are the key reasons for particle loss in a trapped gas, since both types of particles (dimer and atoms) in the final product will have a very high kinetic energies due to energy-momentum conservation, and will leave the trap.  It should be noted  that these two processes can only be activated after a population of shall dimers is formed as a result of Process ${\bf (I)}$ and ${\bf (I')}$. 
 Our study shows that the atom loss rate depend on a variety of factors such as trap depth and the temperature of the system. The location of maximum loss rate is not ``intrinsic" in the sense that it can change by varying external conditions such as trap depth, even though the gas parameter $k_Fa_s$ is kept fixed, where $k_F$ is the Fermi wave vector in the center of the trap.\\

The paper is organized as the following. In Sec.\ref{gd}, we give an overview of the recent experiments and  the range of parameters appeared in these studies. In Sec.\ref{2sp}, we discuss the processes  $({\bf I})$ to ${\bf (III)}$ mentioned above. We show that on general grounds, the loss rate will have a maximum at a magnetic field below Feshbach resonance. 
In Sec.\ref{sec_theory}, we implement these processes into a set of rate equations, and give explicit
expressions for the rate constants in these equations. 
In Sec.\ref{ex}, we show that the results obtained from the rate equations provide a good description for all current experiments which are performed under a variety of external conditions. 
In the concluding Sec.\ref{con}, we further discuss the implication of the ``two step" process (i.e. [{\bf (I)} + {\bf (I')}] and [{\bf (II)} + {\bf (III)}] )
mentioned above.

\section{Overview of recent experiments}
\lb{gd}
The experiments we are going to discuss are performed in a mixture of two-component Fermi gases with density $n$ interacting with an $s$-wave scattering length $a_{s}$.  Interaction strength of the system is characterized by the ``gas parameter" $k_{F}a_{s}$, where $k_{F}= (3\pi^2 n)^{1/3}$ is the Fermi wavevector. 
For positive scattering length, $a_{s}>0$, a two-body system can be either in a scattering state or in a bound state. In the literature, these bound states are sometimes referred to as ``Feshbach molecules", or sometimes simply ``dimers". The binding energy of the dimer is $E_b=\hbar^2/ma_s^2>0$.  When the two-body system is in the scattering state (or in the bound state), it is referred to as in the ``upper branch" (or ``lower branch"). 
The generalization of this terminology to the many-body case (for $a_{s}>0$) is the following. If the quantum gas is in thermal equilibrium, it is referred to as in the ``lower" branch. If the quantum gas is prepared in a  state where dimers are absent, 
such as the case at sufficiently high temperature or sufficiently small $k_{F}a_{s}$, it is referred to as in the ``upper branch". 
By definition, the system in the upper branch is not in its ground state. However, if the rate of production of Feshbach molecules is sufficiently low, the atoms in the system can be regarded in a quasi-equilibrium state during the time when experiments are performed. \\


To our knowledge, all recent experiments on atom loss in a trap are performed in the upper branch.   These experiments include many studies of Grimm's group at Innsbruck \cite{Grimm2003}; the work of Jin's group at JILA \cite{Debbie2004}, the work of Ketterle's group at MIT \cite{Ketterle2002,Ketterle2009}, and Salomon's group at ENS\cite{Salomon2003}. Those experiments were performed under a wide range of conditions, with temperature $T$ spanning over two orders of magnitude, trapping depth $V_0$ over three orders of magnitude, and a wide range of Fermi temperatures $T_{F}$. The parameters of these experiments are collected in Table \ref{tb}.  In all these experiments, a maximum loss rate is found at a magnetic field $B^{\ast}$ below the magnetic field $B_{\infty}$ where Feshbach resonance occurs. The positive scattering length at which the maximum loss takes place will be denoted as  $a_s^{\ast}$. The binding energy of the Feshbach molecule (or dimer) at $a_s^{\ast}$ will be denoted as $E_{b}^{\ast}$. 
\begin{center}
\begin{table}[h]
\caption{Experimental Parameters.}
\begin{tabular}{cccccccccccc}
\hline\hline 
$V_0 (\mu K)$ & $T (\mu K)$ & $T_F (\mu K)$ & $k_Fa_s^*$& $B^*$ (G) & $a_s^*$(\AA) & $E_b^* (\mu K)$& $E_b^*$/T & $E_b^*$/$V_0$ & $V_0$/T & Ref \\
\hline 
7.1 & 0.28 & 1.4 & 2 & 790 & 4318& 0.7 & 2.5 & 0.098 & 25.35 & MIT\cite{Ketterle2009}, Li \\
10 & 0.67 & 0.79 & 0.83 & 201.4 & 990& 1.64 & 2.44 & 0.164 & 14.9 & JILA\cite{Debbie2004}, K \\
175 & 22 & 21 & 1.6 & 680 & 700& 17.88 & 0.81 & 0.102 & 7.95 & MIT\cite{Ketterle2002}, Li \\
350 & 22 &  2.8 &  $ 0.359$ & 644 & 427&64 & 2.9 & 0.183 & 15.9 & Innsbruck\cite{Grimm2003}, Li \\
500 & 30 &  2.8 &  $0.319$ & 636 &380 & 83 & 2.76 & 0.166 & 16.6 & Innsbruck\cite{Grimm2003}, Li \\
1000 & 60 &  2.8 &  $0.286$ & 629 & 342 &104 & 1.73 & 0.104 & 16.6 & Innsbruck\cite{Grimm2003}, Li \\
  \footnotemark[1]   & 2.4 & 6 & 1.2 & 720 & 1207 &7 & 2.9 & & & ENS\cite{Salomon2003}, Li\\
\hline\hline
\footnotetext[1]{Ref.\cite{Salomon2003} stated that the trap depth is of the same order as temperature. The specific value for it was not reported. }
\end{tabular}
\lb{tb}
\end{table}
\end{center}

Different groups have different views on the physical origin of the maximal loss. The Innsbruck group interpreted their data as due to a two-step process, similar to the mechanism presented below, but without detailed formulation and calculations \cite{Grimm2003}.  We shall discuss this picture in detail below.  The JILA group noted that at the maximal loss, the heating of the system also reaches a maximum. However, no specific proposals were made for the origin of the observation.   The ENS group considered the maximum loss as a consequence of the rising three-body recombination rate at small $a_{s}$, and the decreasing binding energy of the Feshbach molecule as $a_s$ approaches infinity at resonance. It is argued that the latter reduces heating (and hence loss rate) because the energy release in the 3-body recombination is small.  However,  no quantitative comparison had been made with experimental data using this picture.  In the earlier experiment of the MIT group, no explanation was given to this maximum loss. In ref.\cite{Ketterle2009}, which is performed at very low temperatures, the maximum loss rate is considered as evidence for Stoner ferromagnetism. Their point is that if the system turns ferromagnetic, different spins will segregate, which will naturally leads to a vanishing three body loss rate. The only problem, however, is that no segregation of up and down spins have ever been observed.  This explanation is very different from all previous pictures, for it makes use of intrinsic property of the ground state rather than specific microscopic scattering processes.  It raises the general question of how reliable it is to use  the 3-body loss as a tool to probe of the nature of the ground state, ferromagnetic or not.  \\

\section{The two-step Process}
\lb{2sp}
Let us first examine the loss channels available for the trapped gas.  For simplicity, we start with
a system consists of only atoms, namely, no shallow dimers are
present. In that case, the only loss channel at time $t=0$ is the
three-body recombination (process ({\bf I})), which we schematically denoted as
\be
A_\uparrow + A_\uparrow + A_\downarrow\to A_\uparrow + D ~~~~~\mbox{Process ({\bf I})}
\lb{3ar}
\ee
and a similar process where up and down spins are interchanged.  Here, $A_\uparrow (A_\downarrow)$ stands for atom with spin up (down) atoms and $D$ is a shallow dimer consists of two opposite spins. We have only considered those
three-body recombinations leading to shallow dimers. In principle,  three-body recombinations  can also lead to deep bound states that  exist in the two-body van der Waals potential. However,  the Franck-Condon factors for these transitions are so much smaller than those for the shallow dimers that they render the transitions to deep bound state negligible in comparison. 
Process ({\bf I}) is an exothermic process. An important point to note is that 
the energy released in the process is of the order the dimer binding energy $E_{b}$ 
which is typically much smaller than the trap depth $V_0$, i.e. $E_b\ll
V_0$ (see Table \ref{tb}). As a result, the shallow dimers ($D$) formed in process ({\bf I}) do not leave the trap. 
We shall see that this has important consequences for all other processes to be discussed. 
In the literature, the rate of process ({\bf I}) is often denoted as $L_3$. This rate has been calculated by Petrov \cite{Petrov2003}.  The result is that $L_3$ is proportional to the $a_s^6$ and depends linearly on the average kinetic energy of the particles. \\

Together with Process ${\bf (I)}$ is the reverse process where a dimer dissociates back to atoms through collisions with other atoms and with dimers, which we denote as 
\be
A_{\uparrow(\downarrow)} + D \to A_{\uparrow(\downarrow)} + A_{\uparrow}+A_{\downarrow}~~~~~\mbox{Process ({\bf I'})}.
\lb{add}
\ee
This will lead to an increase of the number of atom and a decrease of the number of shallow dimers ($D$) in the trap.  Process ${\bf (I')}$ shares the same microscopic matrix element as ({\bf I}). However, the density of states of the initial configuration of these two processes  are entirely different.  For the dissociation process ${\bf (I')}$, it depends crucially on temperature and interaction parameters of the system.  The population of shallow dimers in the trap is a result of the competing effects  ${\bf (I)}$ and ${\bf (I')}$. \\

While both ${\bf (I)}$ and ${\bf (I')}$ change the numbers of atoms  and dimers,  none of these particles are lost from the trap during these processes, since   $E_b\ll V_0$ as mentioned before.   Of course, shallow dimers can also dissociate through dimer-dimer collision, 
\begin{equation}
D+D \to  A_\uparrow+A_\downarrow+A_\uparrow+A_\downarrow ~~~~~ \mbox{Process ({\bf I''})}.
\end{equation}
We shall ignore Process {\bf (I'')} since it has the same microscopic matrix element as its reverse process in which four atoms collide to form two shallow dimers, which is very small. 
While ${\bf (I)}$ and ${\bf (I')}$ do not cause particle loss from the trap, processes leading  to formation of deep bound dimer state will. 
These processes arise from collision between shallow dimers and atoms, (referred to as Process ${\bf (II)}$) or between shallow dimers with each other (referred to as Process ${\bf (III)}$). Schematically, they are represented as 
\be
A_{\uparrow(\downarrow)} + D \to A_{\uparrow(\downarrow)} + D^* ~~~~~\mbox{Process ({\bf II})},
\lb{adr}
\ee
\be
D+D  \to D^*+D^* ~~~~~ \mbox{Process ({\bf III})} \lb{ddr}.
\ee
where $D^{\ast}$ represents the deep bound state. Since the energy of the deep bound state is large and negative,  the atoms and dimers in the final state in Process {\bf (II)} and {\bf (III)} will carry very large kinetic energy and will leave the trap. 
The rates of Processes {\bf (II)} and {\bf (III)}, denoted as  $L_2$ and $L_m$, respectively, have also been calculated by Petrov {\it et al.} \cite{Petrov2005}, who found  $L_2\propto a_s^{-3.3}$ and $L_m\propto a_s^{-2.5}$. \\

{\em The Origin of the maximum in atom loss rate:}  The processes above show that the loss of atoms from a trap proceeds in two-steps. The first is to produce a density of shallow dimers, 
which is the combined effect of the three body-recombination ${\bf (I)}$ and dimer disassociation ${\bf (I'})$.  Once the dimers are formed, Processes ${\bf (II)}$ and ${\bf (III)}$ will be activated and produce atoms and dimers with kinetic energies high enough for them to leave the trap.  As one approaches the resonance from small $k_{F}a_{s}$, the rate of 3-body recombination (process ${\bf (I)}$) increases as $a_{s}^6$ \cite{Petrov2003}. The population of the shallow dimer, and hence the atom loss rate, therefore increases with $a_{s}$.  However, as one gets closer to resonance, $k_{F}a_{s}\rightarrow +\infty$, the binding energy of the shallow dimer $E_b = \hbar^2/ma^{2}_{s}$ decreases rapidly, making the dissociations process ${\bf (I')}$ more and more effective, caused by thermal effects at high temperatures, or by Pauli blocking effects in quantum degenerate regime \cite{Combescot2003}, (see later discussions). 
 At some point, dimer dissociation ${\bf (I')}$ will overwhelm the 3-body recombination process ${\bf (I)}$,  thereby quenches Processes {\bf (II)} and {\bf (III)} and reduces the loss rate.  \\

\section{Theoretical Model}
\lb{sec_theory}
In this section, we express the processes discussed in Section \ref{2sp} in terms of rate equations. Our basic assumption is that there is a well-defined degrees of freedom for the shallow dimers. We shall denote the number of atoms of up and down spins as $n_{\uparrow}$ and $n_{\downarrow}$, and the 
number of atoms and number of dimers are denoted as $n_a$ and $n_m$ respectively, and $n_{a}=n_{\uparrow}+n_{\downarrow}$.
The processes in Section \ref{2sp} imply
\begin{widetext}
\ba
\PD{n_m(t)}{t}&=&\left[ L_3 (n^2_\uparrow(t)n_\downarrow(t)+n_\uparrow(t)n^2_\downarrow(t))-qL_3 n_a(t)n_m(t) \right] -L_2 n_a(t)n_m(t)-2L_m n_m(t)^2\lb{rm1}\\
\PD{n_\uparrow(t)}{t}&=&\left[ -L_3 (n^2_\uparrow(t)n_\downarrow(t)+n_\uparrow(t)n^2_\downarrow(t))+qL_3 n_a(t)n_m(t) \right]-L_2 n_\uparrow(t)n_m(t)
\lb{rau}\\
\PD{n_\downarrow(t)}{t}&=&\left[ -L_3 (n^2_\uparrow(t)n_\downarrow(t)+n_\uparrow(t)n^2_\downarrow(t))+qL_3 n_a(t)n_m(t)\right] -L_2 n_\downarrow(t)n_m(t)  \lb{rad}
\ea
\end{widetext}
The first terms on the right hand side of Eqn.(\ref{rm1}) describes the production of dimers through three-body recombination (Process ({\bf I})) with rate $L_3$. The second term describes the dissociation process ${\bf (I')}$ through atom-dimer collision (hence proportional to $n_{a}n_{m}$). We have parametrized the rate of the process as $L_{3}q$. As we shall see, $q$ depends on temperature $T$, scattering length $a_{s}$, and the total number of particles $n=n_{a} + 2n_{m}$; 
\begin{equation}
q=q(T, a_{s}, n),
\end{equation}
since the ratio between atom and shallow dimers in equilibrium depend on the total particle number $n$. (See discussions later in this section).  
That we group these two terms  ($L_{3}$ and $qL_{3}$) together in a square bracket because they do not contribute to particle loss in the trap.  (See discussions in the previous section).  The third term in  Eqn.(\ref{rm1}) describes the loss of shallow dimer in Process ${\bf (II)}$ through collisions between atom and shallow dimer that leads to production of deeply bound states. The rate of this process is $L_{2}$. The last term in  Eqn.(\ref{rm1}) describes Process ${\bf (III)}$, which describes the loss of shallow dimers due to collisions between them to form deeply bound states. The rate of this process is $L_{m}$. 
Eqn.(\ref{rau}) and (\ref{rad}) can be understood similarity, based on the picture that 
 the dimers produced in Process ${\bf (I)}$ do not leave the system, (since the binding energy of the dimer $E_{b}$ is much less than the typical trap depth shown in Table \ref{tb}). As a result, 
 the shallow dimers produced continue to participate in the 3-body recombination process and dissociation process. \\
 
 With $q$ being a function of $n=n_{a}+2n_{m}=n_{\uparrow} + n_{\downarrow} +2n_{m}$,  Eqn.(\ref{rm1}), (\ref{rau}), and (\ref{rad}) form a close set of equations for $n_{\uparrow}$, $n_{\downarrow}$, and $n_{m}$, and allow us to study the time evolution of these quantities.
For the case of equal spin population, $n_\uparrow=n_\downarrow= n_{a}/2$,  Eqn.(\ref{rm1},\ref{rau},\ref{rad}) can be simplified to 
\begin{widetext}
\ba
\PD{n_m(t)}{t}&=& \left[ \frac{L_3}{4} n_a^3(t)-qL_3 
n_a(t)n_m(t) \right]   -L_2 n_a(t)n_m(t) -2L_m n_m(t)^2
\lb{rm}\\
\PD{n_a(t)}{t}&=&-\left[ \frac{L_3}{2} n^3_a(t) -2qL_3 n_a(t)n_m(t)\right] -L_2 n_a(t)n_m(t).
\lb{ra}
\ea
\end{widetext}
Note that from Eqn.(\ref{rm}) and (\ref{ra}),  the total number of particles $n=n_a+2n_m$ decreases as 
\be
\PD{n}{t}=-2L_2n_a(t)n_m(t)-2L_m n_m^2(t), 
\ee
which is the statement that only Processes ${\bf (II)}$ and ${\bf (III)}$ lead to particle loss. \\

To make use these equations, we need to obtain expressions of the parameters $L_{3}, L_{2}, L_{m}$ and the function $q(T,a_{s},n)$. 
The typical values of $L_3$, $L_2$ and $L_m$ can be obtained from existing experiments. Our
strategy is to fix their values at one scattering length $a_s$ and then use the
scaling relation derived by Petrov et al. \cite{Petrov2005} to get the value at other
scattering length. For $^6$Li, the values of $L_3$ and $L_m$ are given in Jochim {\it et
  al} \cite{Grimm2003}. At field $B=690$G, we have $L_3(B=690G)=1\times 10^{-25}cm^6/s$. Thus
at a different magnetic field $B$,
\ba
L_3(B)=\BK{\frac{a_s(B)}{a_s(B=690G)}}^6L_3(B=690G)
\ea
The value of $L_m$ is estimated to be $5\times 10^{-11}cm^3/s$
in S.Jochim {\it et al.} \cite{Grimm2003} at field $B=546$G. According to the
calculation by Petrov {\it et al.}\cite{Petrov2005}, at other magnetic field $B$, the
rate coefficient is given by
\ba
L_m(B)=\BK{\frac{a_s(B)}{a_s(B=546G)}}^{-2.5} L_2(B=546G).
\ea
We haven't been able to find a precise value for $L_2$ in existing experiments. As commented in ref.\cite{Salomon2003a}, the value of $L_2$ cannot be safely estimated, but a reasonable value can be estimated to be $L_2(B=690G)=1\times 10^{-13}cm^3/s$ and thus at another magnetic field
\ba
L_2(B)=\BK{\frac{a_s(B)}{a_s(B=690G)}}^{-3.3} L_2(B=690G).
\ea
Note that strictly speaking, the above formulae were derived for the regime $k_Fa_s< 1$ \cite{Petrov2003,Petrov2005}.  We shall assume in our subsequent discussion that these expressions continue to give a reasonable approximation to actual rates in the region $k_{F}a_{s}\sim 1$. \\

 
To determine $q$, we  use the fact that in the absence of the $L_2$ and $L_m$ terms, the long time evolution of the equations above should establish chemical equilibrium
between atoms and molecules. In that case, we find
\ba\lb{ameq}
n_{a,eq}^2=4 q n_{m,eq}
\label{q} \ea
where $n_{a,eq}=\lim_{t\to\infty} n_a(t)$ and similarly for $n_{m,eq}$. Eqn.(\ref{q}) shows that the quantity 
$q$ is simply the ratio of atoms and molecules in an equilibrium mixture. 
To estimate $n_{a,eq}$ and $n_{m,eq}$ (and hence $q$), we use the simplest model of non-interacting mixture, with the approximate hamiltonian $K=(h_{a}-\mu_{a} n_{a}) +( h_{m}- \mu_{m}n_{m})$, 
where $h_{a}$ and $h_{m}$ are the hamiltonians of the atoms and shallow dimers respectively, and $\mu_{a}$ and $\mu_{m}$ are their chemical potentials, (see also, for example, 
Kokkelmans {\it et al.}\cite{Kokk2004}, as well as Chin and Grimm\cite{Chin2004}). The energy of the dimer will be denoted as $E_{b}$.  Equilibration between atoms and dimers are ensured through the relation of their chemical potentials
\be\lb{chemeqm}
2\mu_a = -E_b +\mu_m, 
\ee
and that the number of atoms $n_{a}$ and the number of dimers are constraint by the condition
\begin{equation}
n= n_{a, eq} + 2n_{m, eq},
\label{n} 
\end{equation}
The explicit for of $n_{a}$ and $n_{m}$ are 
\ba
n_{a,eq}&=&2\int \frac{d^3{\bf k}}{(2\pi)^3}\frac{1}{e^{\b (k^2/2m-\mu_a)}+1}, \label{na} \\ \lb{numsep}
n_{m.eq}&=&\int \frac{d^3{\bf k}}{(2\pi)^3}\frac{1}{e^{\b (k^2/4m-\mu_m)}-1},
\ea
where $\b=1/k_BT$ is the inverse temperature, and $m$ is the mass of the atom. \\

In vacuum, we have $E_{b}=\hbar^2/ma_{s}$. In a many-body system, as temperatures drops to   
quantum degenerate regime, pair formation is affected by the presence of a Fermi sea, which has taken up some momentum states needed for the pair wave function. To capture the effect of Pauli blocking on formation of bound state, we consider the analog of Cooper pairing in the presence of a Fermi sea at finite temperature. 
Denoting the quantum state of Cooper pair as $|\Psi\rangle = \sum_{\bf k} 
\Psi_{\bf k}|{\bf k} \uparrow, -{\bf k} \downarrow \rangle$, the Schrodinger equation of the pair is 
\begin{equation}
E_{b} \Psi_{\bf k} = 2\epsilon_{\bf k} \Psi_{\bf k} + \frac{u_{o}}{\Omega}\sum_{\bf k'} (1-f_{\bf k'})^2 \Psi_{\bf k'}
\label{Cooper} 
\end{equation}
where $\Psi_{\bf k}$ is the amplitude for presence of a pair, $\epsilon_{\bf k} = \hbar^2/2m$, $\Omega$ is the volume, $f_{\bf k}$ is a Fermi function for the atom, 
\begin{equation}
f_{\bf k} = \frac{1}{e^{\beta(\epsilon_{\bf k} -\mu_{a})} +1}, 
\end{equation}
and $u_{o}$ is bare interaction parameter which is designed to reproduce the low energy scattering amplitude, and is related to the s-wave scattering length  $a_{s}$ as 
\begin{equation}
\frac{m}{4\pi\hbar^2 a_{s}} = \frac{1}{u_{o}} + \frac{1}{\Omega}\sum_{\bf k}\frac{1}{2\epsilon_{k}}.
\end{equation}
The reason for the power $2$ in the Fermi exclusion factor $1-f_{\bf k'}$ is because the scattered state consists of two particles ${\bf k'}$ and $-{\bf k'}$. The solution of Eqn.(\ref{Cooper}) is 
\begin{equation}
\frac{m}{4\pi\hbar^2 a_{s}} = \frac{1}{\Omega}\sum_{\bf k}\left[  \frac{ (1-f_{\bf k})^2}{E_{b} - 2\epsilon_{\bf k}} + \frac{1}{2\epsilon_{\bf k}} \right].
\label{Eb} \end{equation}
Eqn.(\ref{Eb}) gives $E_{b}$ as a function of $T$, $a_{s}$ and $\mu_{a}$. It is easy to see that in 
in the non-degenerate limit, $e^{\mu_{a}/T}\rightarrow 0$, we have $f_{\bf k}\rightarrow 0$, and Eqn.(\ref{Eb}) reduces to the equation for bound state energy in vacuum, which gives the usual result 
$E_{b} = \hbar^2/(ma_{s}^2)$. \\

Eq.(\ref{n}) and (\ref{chemeqm}) now imply 
 \begin{equation} 
n = 2\int \frac{d^3{\bf k}}{(2\pi)^3}\frac{1}{e^{\b (k^2/2m-\mu_a)}+1} + 
2 \int \frac{d^3{\bf k}}{(2\pi)^3}\frac{1}{e^{\b (k^2/4m-2\mu_a-E_{b})}-1}.
\label{neq} \end{equation}
Since $E_{b}$ is a function of $T$, $a_{s}$ and $\mu_{a}$, 
$E_{b}= E_{b}(a_{s}, T, \mu_{a})$. Eq.(\ref{neq}) gives $\mu_{a}$ as a function of $n$, $T$, and $a_{s}$. Once $\mu_{a}= \mu_{a}(n,T, a_{s})$ is determined, we can then calculate $n_{a}$ and $n_{m}$ from Eqn.(\ref{na}) and (\ref{numsep}), and obtain $q$ Eqn(\ref{q}), 
\begin{equation}
q(T, a_{s}, n) = n^{2}_{a,eq}/ (4n_{m,eq}).
\end{equation}
\\

Before proceeding, let us comment on the assumptions of Eqns.(\ref{rm},\ref{ra}). (i) First of all, we have assumed a well defined degrees of freedom for the shallow dimers, which is only valid for small $k_Fa_s$ or high temperature. As a result, our equations will not be accurate very close to the resonance where the dimer degrees of freedom becomes less well defined due to many-body effects. (ii) The most important effect that is not captured by Eqns.(\ref{rm},\ref{ra}) is heating. In principle, there should be an equation of the form $\partial T/\partial t = F(T, n_{a}, n_{m}; a_{s})$, where $F$ describes the effect of re-thermalization of the energy released form various processes. 
We have not attempted to construct this equation. The viewpoint we take is that since the physics of the two-step process has already provided an explanation for the loss maximum qualitatively, (see Section \ref{ex}), it is useful to first find out how well these equations account for current experiments quantitatively so as to determine the validity and usefulness of these rate equations. 
The more elaborate effects of  heating will be explored elsewhere. 
(iii) Throughout our discussion, we shall replace the quantum gas in a harmonic trap with a non-uniform density profile by one in a square box with uniform density. The replacement is mainly for simplicity. While one can perform a full calculation using local density approximation (LDA), for the level of accuracy of our comparison, we believe our simple replacement is sufficient.

\section{Comparison with Experiments} 
\lb{ex}
Essentially, all experiments on atom loss are performed in the following way. One starts with a sample with fixed number of atoms $n$ and no dimers at an initial magnetic field $B_{ini}$ corresponding to a small and positive scattering length $a_s^{ini}$. One then jump to the magnetic field $B_{fin}$, corresponding to the scattering length $a_s$ of interest and wait for a time interval $\D t$. Within this time interval, atoms are converted to shallow dimers as well as escaping from the trap due to formation of deep bound states. At the end of this interval, the system has $n_{a}$ atoms and $n_{m}$ shallow dimers. The number of atoms $n_{a}$ is the quantity of interest.  To image the number of remaining atoms after the interval $\D t$, one pulls back the magnetic field $B$ from $B_{fin}$ to $B_{measure}$,  typically corresponds to a scattering length that is small and positive  $a_s^{measure}$.
The rate of the pull back is such that all the shallow dimers formed at magnetic field $B_{fin}$ are converted to tightly bound state of size $a_s^{measure}$ and therefore will not be counted by the imaging process for atoms. The number of atoms counted at $B_{measure}$ (or at $a_s^{measure}$) therefore gives the number of atoms $n_{a}$ at the end of the time interval $\D t$.\\

We have applied the rate equations in Sec.\ref{sec_theory} to different experiments listed in Table \ref{tb}. The results and the parameters used in our calculations are summarized below: 

\subsection{MIT/2002}
In this experiment \cite{Ketterle2002}, the initial ensemble consists of $3\times 10^5$ $^6$Li atoms with peak density $3\times 10^{13}$cm$^{-3}$. The temperature where the final measurement is made is around $T=22\mu K$, slightly higher than the Fermi temperature $T_F=21\mu K$. The trapping parameters are $\o_r=12 Hz$, $\o_z=200Hz$ and the trap depth $V_0=175 \mu K$. The magnetic field is turned on within $4$ms to $B_{fin}$ and one waits for $50$ ms or $500$ms at $B=B_{fin}$, then the magnetic field is  switched off within $100\mu s$ and the cloud is probed by absorption imaging. The experimental findings are that for incoherent two-component Fermi gas,  there is a strong loss around magnetic field $B=680G$. Also, for $B>680G$ and close to the unitarity $B_0=834G$, the loss decreases and saturates.\\

To describe the experiment, we choose as our initial conditions for the rate equations $n=n_a(t=0)=3\times 10^{13}cm^{-3}$ and $n_m(t=0)=0$, corresponding to the experiment. We run our rate equations for $\Delta t=0.045$s. We have chosen the following
parameters for the experiment. $L_3(B=690G)=1 \times 10^{-24}cm^6/s$,
$L_2(B=690G)=1\times 10^{-13}cm^3/s$ and $L_m(B=546G)=5\times
10^{-11}cm^3/s$.  The calculated fraction of atoms remained is shown in Figure \ref{mit2002}. We note that the atom loss behavior in this experiment is quite insensitive to the  value of $L_m$.

\begin{figure} 
  \begin{center} 
    \includegraphics[width=5.0 in]{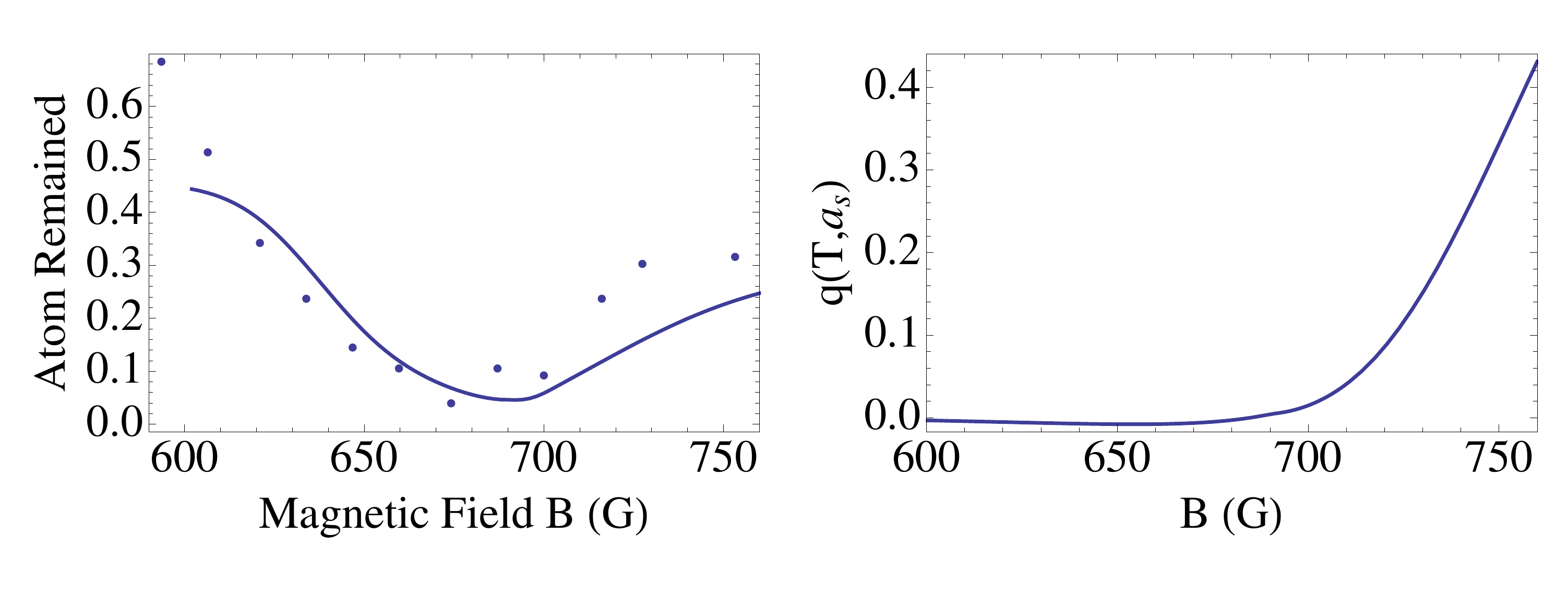} 
    \caption{The calculated atom remaining in the trap for the MIT 2002 experiment. The initial density is given by $n_a(t=0)=3\times 10^{13}cm^{-3}$ and $n_m(t=0)=0$. The corresponding function $q(T,a_s)$ for the appropriate density is shown as well.}
      \lb{mit2002}
 \end{center} 
\end{figure}



\subsection{Innsbruck/2003}
This set of experiments \cite{Grimm2003} are described in detail in S. Jochim's thesis\cite{Grimm2003}. One usually  starts with about two million atoms in the lowest two hyperfine-Zeeman states of the $^6$Li atoms. The samples are cooled to three different temperatures: $22\mu$K, $30\mu$K and $60\mu$K, at a magnetic field $300$G where the scattering length is large and negative. As a result, there are no shallow dimers in the initial state.  One then ramps the system to close to Feshbach Resonance within $50$ms. After waiting for $5\sim 7$s, the magnetic field is ramped back to zero, at which point the number of atoms are measured. The atom loss is found to have a maximum at $636$G.  To describe the experiment, we run the
equation for $\Delta t=5s$, with $n=n_{a}(t=0)= 3.8\times 10^{13}cm^{-3}$, $n_{m}(t=0)=0$ estimated from ref.\cite{Grimm2003}, and $L_3(B=690G)=8\times 10^{-25}cm^6/s$,
$L_2(B=690G)=1\times 10^{-14}cm^3/s$ and $L_m(B=546G)=5\times
10^{-11}cm^3/s$. The calculated number of atoms remained is shown in Figure \ref{inns2004}. \begin{figure}
  \begin{center} 
    \includegraphics[width=5.0 in]{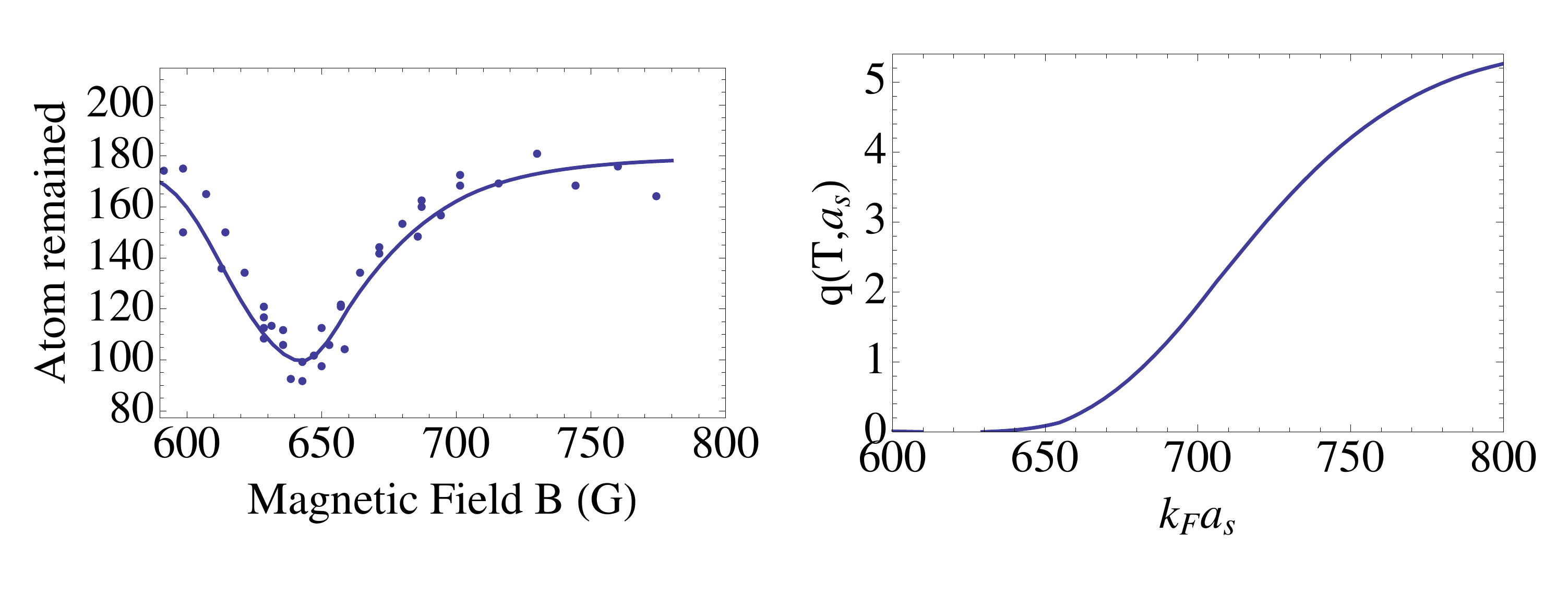} 
    \caption{The calculated atom remaining in the trap for the Innsbruck 2003 experiment. The initial density is given by $n_a(t=0)=3.8\times 10^{13}cm^{-3}$ and $n_m(t=0)=0$. The corresponding function $q(T,a_s)$ for the appropriate density is shown as well.}
      \lb{inns2004}
 \end{center}  
\end{figure}

\subsection{JILA/2004}
The starting point of this experiment \cite{Debbie2004} is an ensemble of $^{40}$K atoms in the hyperfine-Zeeman states $\ket{9/2,-9/2}$ and $\ket{9/2,-7/2}$, with temperature $T=70$nK and $T/T_F=0.22$. The peak density of the system is $1.5\times 10^{13}$ cm$^{-3}$. In this experiment, one uses the radio-frequency spectroscopic to disassociate the molecules in the states $\ket{9/2,-9/2}$ and $\ket{9/2,-7/2}$ to atoms in the hyperfine-Zeeman states $\ket{9/2,-9/2}$ and $\ket{9/2,-5/2}$. By measuring the atoms in the state $\ket{9/2,-5/2}$, one infers the molecule numbers of the system. \\

The atom loss shows the expected non-monotonic behavior for both spin component. To study this   experimental situation, we choose the temperature of the system to be $0.67\mu K$, and run our rate equation for  $\Delta t=95$ms, which is the holding time in the experiment. The maximum loss occurs at around $201.5$G. The following
parameters are assumed for the experiment. $L_3(B=200G)=1.5 \times 10^{-25}cm^6/s$,
$L_2(B=200G)=2\times 10^{-12}cm^3/s$ and $L_m(B=200G)=1\times
10^{-11}cm^3/s$. We have taken $n=n_{a}(t=0)= 1.5\times 10^{13} cm^{-3}$, $n_{m}(t=0)=0$.  The calculated atom loss rate is shown in Figure \ref{jila2004}. 
\begin{figure}[h] 
  \begin{center} 
    \includegraphics[width=5.0 in]{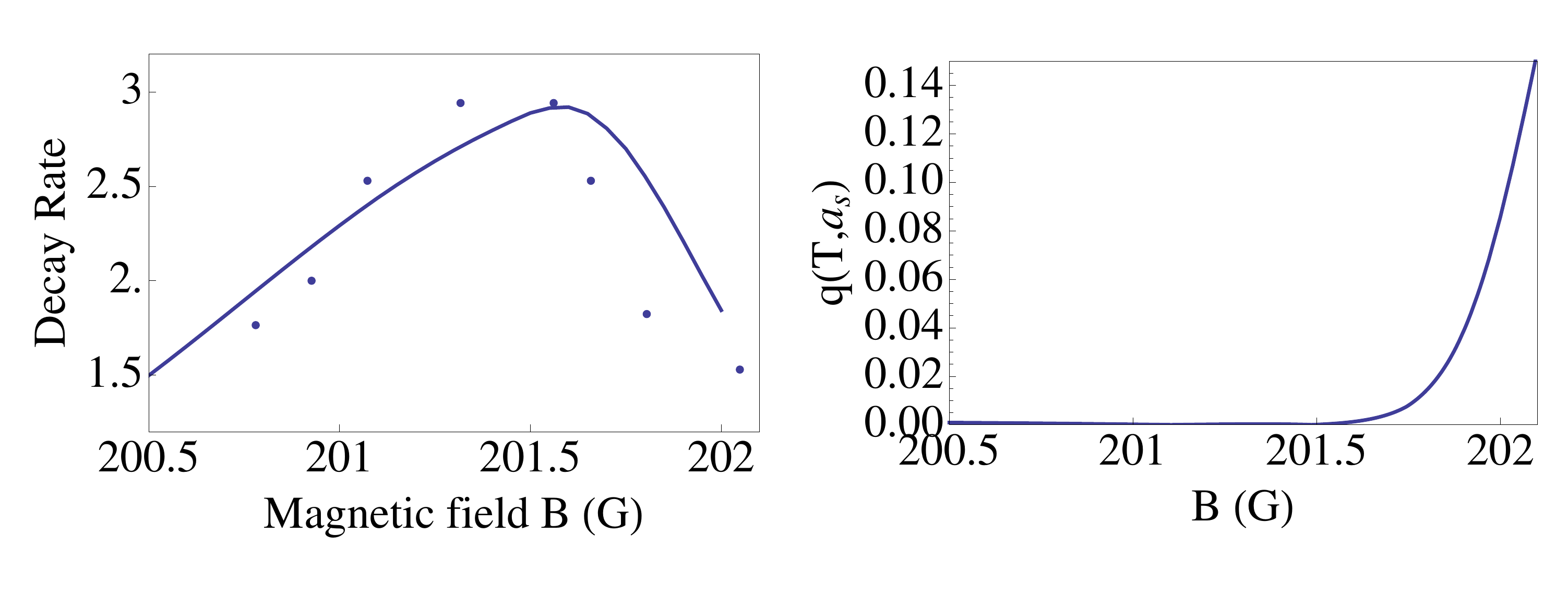} 
    \caption{The calculated decay rate for the JILA 2004 experiment. The initial density is given by $n_a(t=0)=1.5\times 10^{13}cm^{-3}$ and $n_m(t=0)=0$. The corresponding function $q(T,a_s)$ for the appropriate density is shown as well.}
    \lb{jila2004}
 \end{center} 
\end{figure}

\subsection{MIT/2009}
In this experiment \cite{Ketterle2009}, the Fermi gas is prepared initially  at a field $B$ around $600$G. The central density of the trapped gas is $n(r=0)= 0.69\times 10^{13} cm^{-3}$. The trap frequencies are $\nu_x=\nu_y=300$Hz and $\nu_z=70$Hz .  In this experiment, the field is ramped down within $4.6$ms to the desired final field. It is noted that for $k_Fa_s>1.8$,
there is approximately 25\% percent molecule population after the ramp, independent of temperature (within the range considered). A maximum in atom loss similar to the observations of other groups is found, but occurs around $780$G. From the information in Ref.\cite{Ketterle2009}, we take $T=0.3\mu K$.   To study this experiment using our rate equation, we choose  $L_3(B=690G)=1 \times 10^{-29}cm^6/s$,
$L_2(B=690G)=1\times 10^{-13}cm^3/s$ and $L_m(B=546G)=1\times
10^{-11}cm^3/s$. We run our equation for $\Delta t=2$ms, since it was remarked in Ref.\cite{Ketterle2009} that the rate are ``measured within the first 2 ms". We have run our equations with three different initial densities $n=n_{a}(t=0)$ chosen as follows. Since we have replace the trapped gas in a harmonic well by one in a cubic box, we take $n$ as some average  density of the density in the harmonic trap, i.e. setting $n=n_{TF}(r^{\ast})$, where $n_{TF}(r)$ is the Thomas-Fermi density profile of the Fermi gas in the harmonic trap, and $r^{\ast}$ is some radius less than the Thomas-Fermi radius $R_{TF}$ ($r^{\ast}<R_{TF}$). 
The calculated atom loss rate is shown in Figure \ref{mit2009}.  
The three curves in  Figure \ref{mit2009} are different choices of $n$ corresponding to 
different choses of $r^{\ast}$.  The color scheme in 
Figure \ref{mit2009} is that blue, purple, and brown correspond to $r^*=\sqrt{2/5}R_{TF}$, $r^*=\sqrt{1/2}R_{TF}$, and $r^*=\sqrt{3/5}R_{TF}$, respectively. 

\begin{figure}
  \begin{center} 
    \includegraphics[width=5.0 in]{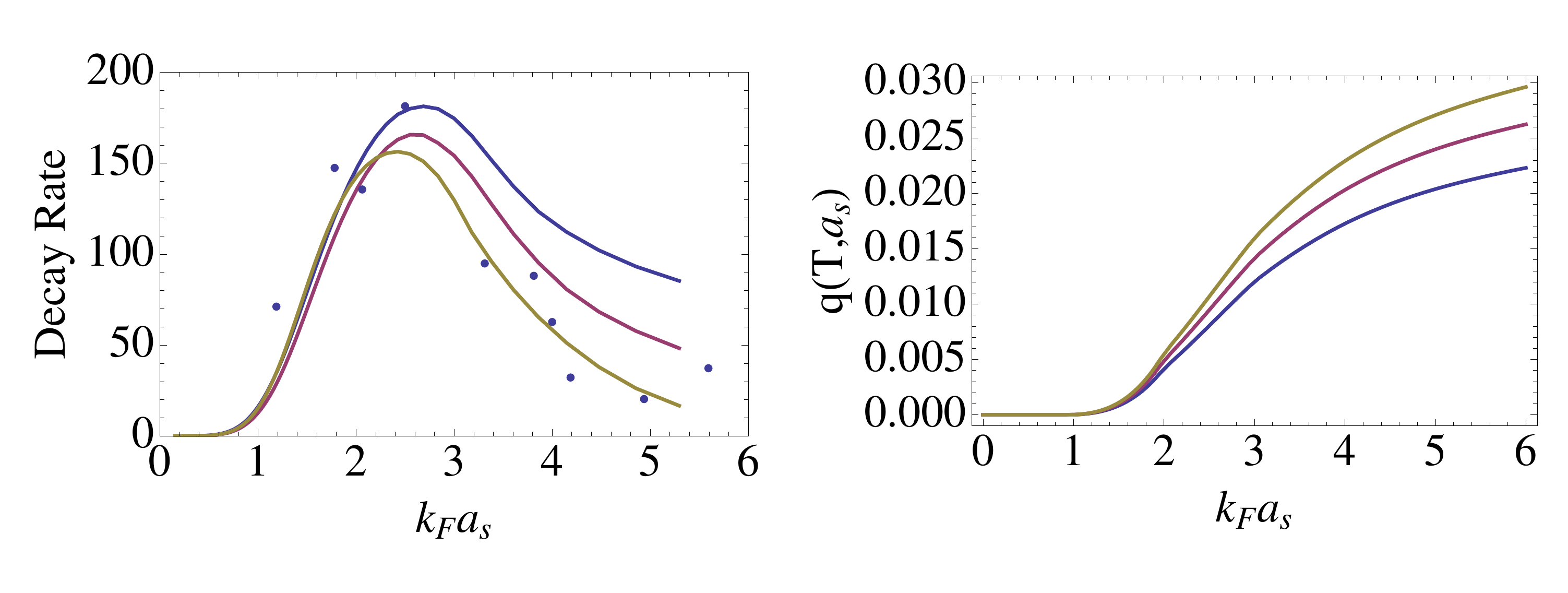} 
    \caption{Calculated decay rate for the MIT 2009 experiment. The initial atomic density is about $n_a(t=0)=0.69\times 10^{13} cm^{-3}$. The three curves correspond to different choice of the positions in the trap. Blue: $r^*=\sqrt{2/5}R_{TF}$; purple: $r^*=\sqrt{1/2}R_{TF}$ and brown: $r^*=\sqrt{3/5}R_{TF}$. The corresponding curve for $q(T,a_s)$ is shown as well.}
    \lb{mit2009}
 \end{center} 
\end{figure}

 \subsection{ENS/2004}
This experiment \cite{Salomon2003} starts with a gas of $7\times 10^4$ $^6$Li atoms in anisotropic trap with $\omega_x=2\pi\times 0.78$kHz, $\omega_y=2\pi\times 2.1$kHz and $\omega_z=2\pi\times 2.25$kHz. The temperature of the system is estimated to be $T=2.4\mu$K, while the Fermi temperature $T_F\approx 6\mu$K. The system is prepared at low magnetic field and then evaporative cooling is performed at field $B=320$G, where the scattering length $a_s=-8$nm. In 10ms, the magnetic field is ramped to anywhere between $600$ and $850$G and time-of-flight expansions are taken and number of atoms is counted. It is found that the loss rate has a maximal around $B=720$G. Interestingly, together with this loss maximum, the sign of interaction energy changes at the same point. \\

The following numbers are used to fit the experiment. The central density is given by $n({\bf r}=0)=3.5\times 10^{13}cm^{-3}$. $L_3(B=690G)=1.5 \times 10^{-26}cm^6/s$,
$L_2(B=690G)=4\times 10^{-13}cm^3/s$ and $L_m(B=546G)=1\times
10^{-12}cm^3/s$. We run the equation for $\D t=1.5$s. We have not been able to find the holding time in ref.\cite{Salomon2003}. However, it is remarked in the cited reference that the life time of the gases ranges from $100 ms$ to few seconds. Our choice of $1.5$s is a rough estimated of the holding time. We have checked that a different choice of reasonable holding time does not change the loss behavior, provided that we modify appropriately other parameters in the calculation.  The calculated atom loss rate is shown in Figure \ref{ens2004}.\\
 \begin{figure}
 \begin{center}
   \includegraphics[width=5.0 in]{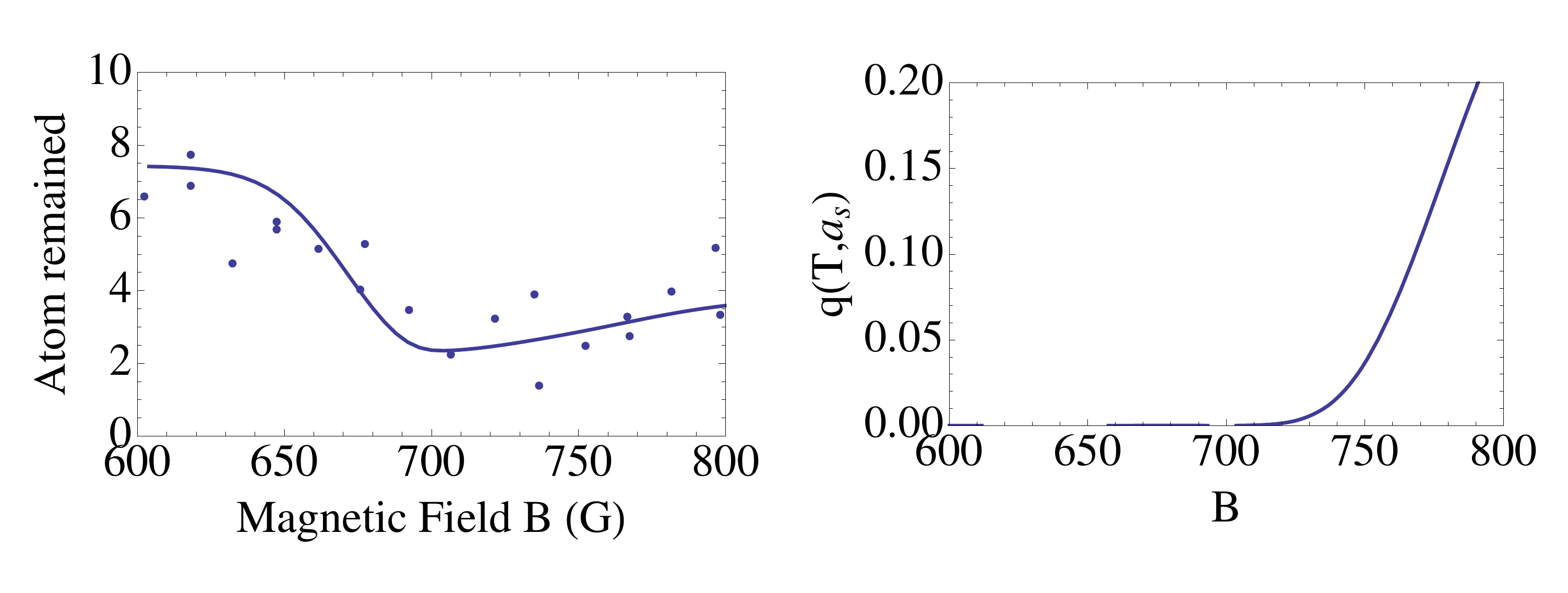}
   \caption{The calculated atom remaining in the trap for the ENS 2004 experiment. The initial density is given by $n_a(t=0)=3.5\times 10^{13}cm^{-3}$ and $n_m(t=0)=0$. The corresponding function $q(T,a_s)$ for appropriate density is given by as well.}
   \lb{ens2004}
 \end{center}
\end{figure}

Finally, we would like to point out that in all our studies, the emergence of a maximum atom loss as a function of $a_{s}$ is a robust phenomena. Modest changes of input parameters, as well as the rates $L_{2}, L_{3}$ and $L_{m}$ do not change our results. 

\section{conclusions}
\lb{con}

In this work, we investigate the origin of the maximum of atom loss as a function of scattering length observed in many experiments that are performed over a wide range of temperature, trap depths, and particle numbers.  We find that this is a result of a two-step process.  The first is the production of a population of shallow dimers which remains in the trap. The second is the conversion of these shallow dimers into deep bound states through collision processes.  The second step, which  causes  particle loss from the trap, depends on the number of shallow dimers generated in the first step.  The maximum of atom loss is caused by the variation of the number of shallow dimers  as a function of $a_{s}$. 
This number is the result of the competition between 3-body recombination and the dimer dissociation processes.  
While the former increases the dimer population at a rate rises as $a_{s}^6$, the later reduces it  and becomes more and more  effective as one approaches the resonance, since the binding energy of the dimer decreases rapidly.  
We have cast these processes in a set of simple rate equation, and show that they can account for the atom loss observed in all current experiments.  \\

Apart from these agreements, we note from  Table \ref{tb} that $E_{b}^{\ast}/V_{o}\sim 0.1$ for all current experiments.  We believe this is not an accident.  Typically, because of the evaporation process, the temperature of the system is lowered if the trapped depth is lowered. Let  $a_{s}^{\ast}$ be the scattering length of the maximum atom loss at temperature $T$. By raising the temperature (caused by a higher depth $V_{o}$), one makes the dissociation process more effective. As a result, one will have to go to a small scattering length (hence larger binding energy $E_{b}^{\ast} = \hbar^2/ma_{s}^2$) to achieve the same ratio between the rate three body recombination and the rate of dissociation. This shows that the observed maximum of atom loss is not purely a function of $k_{F}a_{s}$, but depends on external parameters like trap depth and temperature.  In other words, 
even though experiments are  performed with the same atom density $k_{F}$, the location of the maximum loss will be different for different runs with different trap depth $V_{o}$, and the result would appear irreproducible if variations of extrinsic factors are not fully taken into account. 
In any case, our results show that the maximum of the atom loss can not be used as a tool to determine the nature of the ground state of the system. \\

This work is supported by NSF Grant DMR-0907366 and by DARPA under the Army Research Office Grant Nos. W911NF-07-1-0464, W911NF0710576.

\end{document}